\documentclass{article}
\usepackage{spconf,amsmath,graphicx,xcolor, amssymb, multirow, float, hyperref}

\title{Fine-Grained Engine Fault Sound Event Detection \\ Using Multimodal Signals}
\name{Dennis Fedorishin$^{1}$, Livio Forte III$^{2}$, Philip Schneider$^{2}$, Srirangaraj Setlur$^{1}$, Venu Govindaraju$^{1}$}
\address{$^1$University at Buffalo, Center for Unified Biometrics and Sensors, $^2$ACV Auctions
}

\begin{document}

\ninept
\maketitle

\begin{abstract}

Sound event detection (SED) is an active area of audio research that aims to detect the temporal occurrence of sounds. In this paper, we apply SED to engine fault detection by introducing a multimodal SED framework that detects fine-grained engine faults of automobile engines using audio and accelerometer-recorded vibration. We first introduce the problem of engine fault SED on a dataset collected from a large variety of vehicles with expertly-labeled engine fault sound events. Next, we propose a SED model to temporally detect ten fine-grained engine faults that occur within vehicle engines and further explore a pretraining strategy using a large-scale weakly-labeled engine fault dataset. Through multiple evaluations, we show our proposed framework is able to effectively detect engine fault sound events. Finally, we investigate the interaction and characteristics of each modality and show that fusing features from audio and vibration improves overall engine fault SED capabilities. 

\end{abstract}
\begin{keywords}
Sound event detection, Engine fault detection

\end{keywords}
\vspace{-0.2em}
\section{Introduction}
\label{sec:intro}
\vspace{-0.5em}

Automobile engines are highly-complex mechanical systems that require consistent maintenance for normal operation. Occasionally, engines may develop faults, often through broken or worn components. These faults are often subtle issues that require expert mechanics to diagnose and fix the fault. Expert mechanics often use \textit{sound} and \textit{vibration} to diagnose vehicle engines, as engine faults often emit unique sound and vibration characteristics, for example, metal-on-metal knocking of broken components, or excessive vibration from engine misfires \cite{denton2006advanced}. 

As a result, automatic engine fault detection and machine condition monitoring have become active areas of research that use these signals to try to automatically monitor and diagnose mechanical faults \cite{henriquez2013review}. Works like \cite{adaileh2013engine} use signal processing techniques on engine audio recordings to diagnose faults. Similarly, \cite{tao2019intelligent} use these techniques on accelerometer-recorded vibration instead of audio. Works like \cite{delvecchio2018vibro} explore both of these modalities together to explore differences in using audio and vibration to detect engine faults. More recently, deep learning has been successfully applied to automatic engine fault detection, using both audio \cite{kdd, shahid2022real} and vibration signals \cite{kdd, junior2022fault}. \cite{kdd} recently proposed a large-scale multimodal engine fault detection framework that performs sample-level classification of broad engine faults, across a wide variety of vehicles.

However, these works perform sample-level classification of engine faults, which only give a high-level understanding of an engine's condition. In this work, we seek to extend engine fault detection into \textit{sound event detection} (SED), which is the task of detecting the temporal occurrence of sound events, with onset and offset times. Specifically detecting engine fault sound events at this granularity gives greater insight into present faults, as not only is the occurrence of a fault being detected, but the timing and duration of the fault as well. For example, a similar-sounding engine fault occurring at the startup of an engine may give clue to different faults compared to a similar sound when an engine is idling. Similarly, short-duration abnormal sounds may indicate other faults than sounds that are present for long durations \cite{denton2006advanced}. Extending previous works by understanding what faults are occurring and \textit{when} they occur significantly increases the informativeness of automatic engine fault detection systems.

SED is an active area of research with multiple works spanning application in detecting domestic and urban sounds, and others \cite{turpault2019sound, salamon2017scaper, mesaros2016tut}. Many works focus on improving deep learning architectures for SED, including the convolutional recurrent neural network (CRNN) and its variations \cite{jiakai2018mean,de2021multi,guan2023subband,ebbers2021forward}, and transformer-based architectures \cite{miyazaki2020weakly,li2023ast}. Others focus on developing new loss functions \cite{kothinti2022temporal} and postprocessing strategies \cite{dinkel2021towards}. Given the high cost of creating strongly-labeled sound events, others explore performing SED in weakly-labeled and semi-supervised learning settings. Recent works have developed new strategies for weakly-labeled SED including new architectures \cite{ebbers2021forward, miyazaki2020weakly} and training strategies \cite{turpault2019sound, dinkel2021towards}. Similarly, works like \cite{turpault2019sound,de2021multi,ebbers2021forward,serizel2018large} improve SED with unlabeled data, by leveraging semi-supervised learning methods like the mean teacher algorithm \cite{tarvainen2017mean}. Additionally, the annual DCASE Task4 Challenge \cite{turpault2019sound} focuses on weakly-labeled and semi-supervised SED in domestic environments. In this paper, we draw upon these works and apply them onto fine-grained engine fault SED. Overall, our contributions are: 1) we collect a strongly-labeled dataset of ten fine-grained engine fault sound events across a wide variety of vehicles, 2) we propose a multimodal fusion SED model that predicts engine fault sound events using audio and accelerometer-recorded vibration, and 3) we introduce a pretraining scheme to overcome our limited-size dataset by pretraining on a weakly supervised engine fault dataset.  

\begin{figure*}
  \centering
  \includegraphics[width=1.0\textwidth]{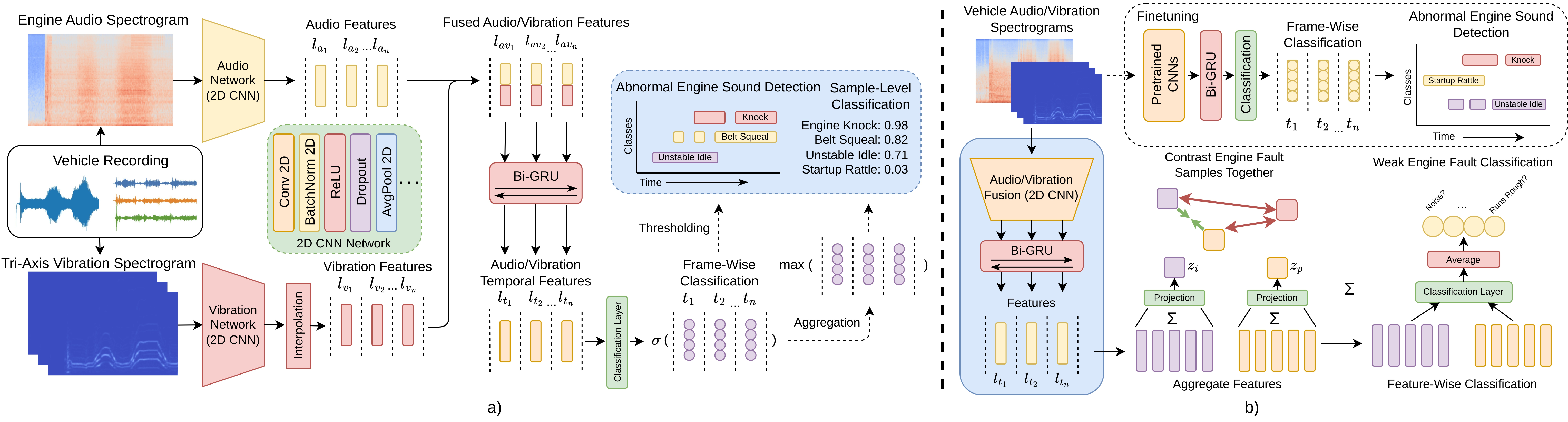}
  \vspace{-2.5em}
  \caption{a) Proposed engine fault sound event detection architecture using audio and vibration engine recordings. b) Proposed pretraining strategy using a large-scale weakly-labeled engine fault dataset with our supervised contrastive loss. Best viewed with zoom and color.}
  \label{fig:model}
  \vspace{-1.2em}
\end{figure*}

\vspace{-1.0em}
\section{Method}
\label{sec:method}

\vspace{-1em}
\subsection{Dataset}
\label{ssec:dataset}
\vspace{-0.5em}

To perform engine fault sound event detection, we collect a dataset of a large variety of common vehicles used in the United States, with the collaboration of professional vehicle condition inspectors of ACV Auctions, an online automotive marketplace. For every vehicle, we collect a 25-35 second audio and vibration recording using a professional-grade microphone and tri-axial accelerometer co-located on the same device, placed inside the vehicle engine bay. When recording, the vehicle is initially off, then turned on with an idle period, and finally accelerated 2-3 times to ensure each state of the engine is captured in the recording. The recorded audio and vibration are temporally consistent, with equal start and stop times.

Given the nature of vehicle engines, many engine faults are often subtle and difficult to detect. As a result, we leverage automotive engine experts that are asked label engine fault sound events of each vehicle, given all of the recorded information. As shown in Table \ref{tab:dataset_details}, we label ten fine-grained engine faults that are broad enough to encompass most engine types, while being specific enough give fine-grained insights into engine condition. Engine knocks, ticks, and unstable idle are often more serious faults that occur deep within the engine internals. Belt squeal and accessory, exhaust, and rattling noises are often audible from non-critical accessory components attached to the engine. Startup rattle and trouble starting are two faults that occur specifically during the startup of an engine. These faults encompass a wide variety of common issues with an engine that when correctly detected, gives direct insight into the engine's condition and repairs necessary to resolve it. 

Table \ref{tab:dataset_details} shows our labeled dataset across the ten engine faults. As shown, most sound events having an average time of over 5 seconds, while certain events like startup rattle are often quickly-occurring that last about 1 second. The entire dataset consists of over 5,000 sound events across 2,643 audio-vibration samples, spread across 232 unique vehicle models. We create a train/validation/test split by splitting the samples such that there is a 70\%/15\%/15\% distribution of each class of sound events across the sets.

\begin{table}[]
\centering
\begin{tabular}{lcccc}
\hline
 \multirow{2}{*}{Class}                 & \multirow{2}{2cm}{\centering Avg. Sound Event Time (s)}& \multicolumn{3}{c}{\# Events} \\ 
           &              & Train    & Valid.   & Test   \\ \hline
Engine Knock      & 9.07         & 531      & 97       & 117    \\
Belt Squeal      & 12.00         & 93       & 19       & 21     \\
Exhaust Noise    & 9.66          & 460      & 86       & 102    \\
Unstable Idle    & 4.40          & 130      & 24       & 29     \\
Internal Tick    & 10.48         & 211      & 39       & 46     \\
Ambiguous Tick   & 9.58          & 381      & 70       & 83     \\
Accessory Noise  & 6.82          & 735      & 136      & 155    \\
Engine Rattle    & 5.25          & 807      & 145      & 168    \\
Startup Rattle   & 1.21          & 199      & 37       & 42     \\
Trouble Starting & 2.09          & 234      & 42       & 49     \\ \hline
\end{tabular}
\vspace{-0.5em}
\caption{Engine fault sound event detection dataset overview.}
\label{tab:dataset_details}
\vspace{-1.5em}
\end{table}

\begin{table*}
\centering
\begin{tabular}{lcccccc|cc}
\hline
                                   & Setup                                               & PSDS$_1$         & PSDS$_2$        & PSDS$_3$         & EB-F1          & SB-F1          & mROC           & mAP            \\ \hline
Random                             & \multicolumn{1}{c}{\multirow{4}{*}{No Pretraining}} & .0208          & .0008         & .0013          & .0053          & .0629          & .5031          & .1033          \\
Audio Only                         & \multicolumn{1}{c}{}                                & .5036          & .3968         & .2876          & .0979          & .3449          & .7586          & .3384          \\
Vibration Only                     & \multicolumn{1}{c}{}                                & .3690          & .2272         & .1958          & .0385          & .1636          & .6249          & .1853          \\
\textbf{Audio + Vibration}         & \multicolumn{1}{c}{}                                & \textbf{.5207} & \textbf{.4024}& \textbf{.3289} & \textbf{.1020} & \textbf{.3579} & \textbf{.7646} & \textbf{.3532} \\ \hline
\multirow{6}{3cm}{Audio + Vibration (Pretrain + Finetune)}&$\lambda_1=1.0, \lambda_2=0.0$& .5346          & .4078         & .3492          & .0980          & .3753          & .7758          & .3890          \\
                                   & $\lambda_1=0.0, \lambda_2=1.0$                      & .5224          & .4130         & .3296          & .0934          & .3380          & .7579          & .3349            \\
                                   & $\lambda_1=1.0, \lambda_2=0.2$                      & .5458          & .425          & .3698          & .1010          & .3758          & .7790          & .3759          \\
                                   & $\boldsymbol{\lambda_1=1.0, \lambda_2=0.5}$         & .5524          & .4315         & \textbf{.3799} & \textbf{.1163} & \textbf{.4067} & .7761          & \textbf{.3966} \\
                                   & $\lambda_1=1.0, \lambda_2=1.0$                      & \textbf{.5618} & \textbf{.439} & .3760          & .1046          & .3882          & .7849          & .3795          \\
                                   & $\lambda_1=1.0, \lambda_2=2.0$                      & .5588          & .4319         & .3752          & .1050          & .3773          & \textbf{.7870} & .3653          \\ \hline
\end{tabular}

\vspace{-0.5em}
\caption{Quantitative results on engine fault sound event detection. Each result is the average of three models with random initializations.}
\vspace{-1.5em}
\label{tab:results}

\end{table*}

\vspace{-1em}
\subsection{Model Architecture}
\label{ssec:modelarch}

Our proposed model, shown in Figure \ref{fig:model}a, is a multimodal fusion SED model built upon the CRNN architecture, widely used across other sound event detection applications \cite{jiakai2018mean, guan2023subband, ebbers2021forward}. We extract features from the audio and vibration spectrograms using independent CNN networks, then fuse them and pass the fused features into a bi-directional GRU network. Finally, we perform a frame-wise classification that yields the final SED predictions across time steps, which are thresholded and aggregated to yield an event-based temporal detection and sample-level classification, respectively. 

Given a vehicle that has a recorded audio and vibration sample, we construct a log-Mel spectrogram $X_a \in \mathbb{R}^{T_a \times F_a}$ of the audio and a linear magnitude spectrogram $X_v \in \mathbb{R}^{3 \times T_v \times F_v}$ of each of the three accelerometer directions. $T_a \times F_a$ and $T_v \times F_v$ denote the number of frames and frequency bins of the audio and vibration representations, respectively. The audio CNN, denoted by $f_a(X_a)$, extracts features from the audio spectrogram $X_a$, resulting in a frame-wise feature representation $l_a \in \mathbb{R}^{t_a \times z_a}$, where $t_a$ and $z_a$ denote the number of time steps (frames) and feature vector size, respectively. $f_a$ consists of seven repeated blocks comprised of a 2D convolution, batch normalization, ReLU activation, dropout, and an average pooling layer. Similar to the audio CNN, the vibration CNN, $f_v(X_v)$, extracts features from the tri-axis vibration spectrogram $X_v$, resulting in a frame-wise feature representation $l_v \in \mathbb{R}^{t_v \times z_v}$, with $t_v$ and $z_v$ denoting the number of vibration time steps and feature size, respectively. $f_v$ consists of six repeated blocks with the same structure as $f_a$. Since each modality has different-sized spectrogram representations, the resulting representations $l_a$ and $l_v$ have the same channel dimension, but different time step amounts. To match the number of time steps $t_a$ and $t_v$, we use nearest-neighbor interpolation across time steps to match the vibration features to audio, such that $t_v = t_a$. 

Next, we fuse the audio and vibration features by concatenating features across time steps, resulting in $l_{av} = [l_a, l_v] \in \mathbb{R}^{t \times (z_a + z_v)}$. The fused features $l_{av}$ are then passed through a bi-directional GRU, denoted by $f_{GRU}(l_{av})$, to extract temporal features and dependencies between each modality and time steps, resulting in $l_t \in \mathbb{R}^{t_{GRU} \times z_{GRU}}$, where $z_{GRU}$ and $t_{GRU}$ is the resulting hidden state size and number of time steps, respectively. Finally, we perform a frame-wise classification using a linear layer, $f_C$, and sigmoid activation, $\sigma$, resulting in the final output of the model $\hat{y}_s \in \mathbb{R}^{n_t \times C}$, where $n_t$ and $C$ denote the final number of time steps and classes, respectively. The overall model $f$ is written as: 
\begin{equation}
\hat{y}_s = f(X_a,X_v) = \sigma (f_C(f_{GRU}([f_a(X_a), f_v(X_v)])))
\label{eq:model}
\end{equation}
To create a sample-level classification alongside $\hat{y}_s$, we simply take the maximum prediction of each class across time steps, denoted by $\hat{y}_w = \max_{n_t}(\hat{y}_s) \in \mathbb{R}^{C}$.

\vspace{-0.5em}
\subsection{Implementation Details}
\label{ssec:implementdetails}

The audio and vibration samples are zero padded and cropped to 30-second signals, with a sample rate of 44.1kHz and 416Hz, respectively. The audio spectrogram $X_a$ is a log-scaled Mel-spectrogram using 128 Mel bins, and a frame size and hop length of 2048 and 1024 samples, respectively. The vibration spectrogram $X_v$ is a linearly-scaled magnitude spectrogram using a frame size and hop length of 256 and 32 samples respectively, with 129 frequency bins. For $X_a$ and $X_v$, we perform channel-wise Z-score normalization. Each convolution layer in $f$ uses a kernel size of $(3\times 3)$. For $f_a$, we use an average pooling kernel of $(2\times 2)$ of three blocks and $(1\times 2)$ for the remaining blocks. Similarly for $f_v$, $(2\times 2)$ is used for one block and $(1\times 2)$ for the remaining. For both $f_a$ and $f_v$, the blocks have channel sizes of 16, 32, 64, and 128 for the remaining blocks, respectively. For $f_{GRU}$, we use a hidden state size of 128. For all dropout layers, we set $p=0.5$. We use 161 time steps for the audio, vibration, and GRU network, which results in a resolution of about 0.2 seconds per time step. The model is trained for 100 epochs using binary cross entropy loss with the AdamW \cite{adamw} optimizer, with a learning rate of 0.001, weight decay of 0.02, and batch size of 48. 

\vspace{-1.0em}
\subsection{Large-Scale Pretraining}
\label{ssec:pretraining}

Using weakly-supervised training to improve SED has been shown to be successful, lessening the need of large amounts of strong labels \cite{turpault2019sound, ebbers2021forward, serizel2018large}. For engine faults specifically, labeling sound events is extremely expensive, often involving multiple engine experts to discern subtle faults. Therefore, as shown in Figure \ref{fig:model}b, we utilize an existing weakly-labeled engine fault dataset to \textit{pretrain} our proposed SED model. Our hypothesis is that pretraining the model to perform sample-level classification of broad engine faults will provide a strong initialization when training for fine-grained engine fault SED.

To do so, we utilize the large-scale multimodal engine fault dataset from \cite{kdd}, which contains over 100k audio and vibration recordings of vehicles with sample-level labels for five multi-label broad engine faults. To pretrain, we utilize a combination of a multilabel classification loss and a supervised contrastive loss. In literature, contrastive losses have been shown to create strong and discriminative embedding spaces for a wide variety of tasks \cite{le2020contrastive}. When labels are present, simple classification losses, supervised contrastive loss \cite{supcon}, and combination of them \cite{dao2021multi} have been shown to be strong learning objectives. 

Since only sample-level labels are available for pretraining, we average the output $\hat{y}_s$ of model $f$ across time steps, denoted by $\hat{y} = \frac{1}{n_t}\sum_{i\in n_t}\hat{y}_{s_i} \in \mathbb{R}^C$, yielding a sample-level output used in the classification loss. Note we average across time steps rather than the max operation in 3.2 for better gradient flow across time steps. Similarly, we also average the feature representations $l_t$ before the classification layer, yielding $\overline{l} = \frac{1}{t_{GRU}}\sum_{i \in t_{GRU}} l_{t_i} \in \mathbb{R}^{z_{GRU}}$.  We pass $\overline{l}$ through a projection layer of two linear layers and a ReLU activation, denoted by $f_{proj}$, yielding a projected representation $z$, which are used in the contrastive loss. Similar to \cite{dao2021multi}, we extend supervised contrastive loss \cite{supcon} to the multilabel setting by averaging multiple losses across each class. For a given sample $i$, we define our loss:
\vspace{-1.5em}
\begin{multline}
\mathcal{L}_i = \lambda_1\left(-\frac{1}{C}\sum_{c=1}^{C}y_{i_c}\log(\hat{y}_{i_c})+(1-y_{i_c})\log(1-\hat{y}_{i_c})\right) \\ 
+ \lambda_2 \left(-\frac{1}{C}\sum_{c=1}^{C} \frac{1}{|P(i_c)|}\sum_{p\in P(i_c)}\frac{\text{exp}(z_i \cdot z_p / \tau)}{\sum_{a\in A(i)} \text{exp}(z_i \cdot z_a / \tau)}\right)
\label{eq:pretrain}
\end{multline}
Here, $P(i_c)=\{p\in A(i)|y_{p_c}=y_{i_c}=1\}$, which is the set of all samples in a batch with the same positive class $c$. The loss for an entire batch is the average across all samples $i$, $\mathcal{L}(N) = \frac{1}{N}\sum_{i=1}^{N}\mathcal{L}_i$. The classification term in $\mathcal{L}$ focuses on the classifying engine faults of individual samples, while the supervised contrastive term enforces the similarity of features \textit{across} samples with the same engine faults, creating a more separated and discriminative embedding space.

We pretrain the model $f$ with (\ref{eq:pretrain}) using 100k samples from \cite{kdd}, for 40 epochs. We set $\lambda_1{=}1.0$, $\lambda_2{=}0.5$, and $\tau{=}0.07$, and use the same details in \ref{ssec:implementdetails}. The audio and vibration samples are recorded at 44.1kHz and 100Hz, respectively, with a length of 30 seconds. For vibration, we upsample the signals from 100Hz to 416Hz to match the dataset we collected.  After pretraining, we use the weights of each pretrained CNN $f_a$ and $f_v$, discard $f_{C}$, $f_{proj}$, $f_{GRU}$, and finetune the model $f$ using the same details in \ref{ssec:implementdetails}. We found that discarding the pretrained weights of $f_{GRU}$ yields better finetuning performance. We hypothesize that the pretrained $f_{GRU}$ is not useful for SED as there is no temporal information from the weak labels.

\vspace{-1.0em}
\section{Experiments}
\label{sec:experiments}
\vspace{-0.5em}

\subsection{Evaluation Metrics}
\label{ssec:evaluation_metrics}
To evaluate engine fault detection performance, we follow standard metrics used in SED. Specifically, we use the Polyphonic Sound Detection Score (PSDS) \cite{psds, psds_eval} under multiple settings, and segment- and event-based F1 scores \cite{sedmetrics}.  For PSDS, we use three settings, denoted by PSDS$_{1..3}$. For PSDS$_1$ and PSDS$_2$, we set $\rho_{DTC}$,$\rho_{GTC}$,$\alpha_{ST}$ = (.05,.05,0) and (0.4,0.4,0), which evaluate detection performance with relaxed and strict intersection tolerances, respectively. For PSDS$_3$, we set $\rho_{DTC}$,$\rho_{GTC}$,$\alpha_{ST}$ = (.05,.05,1.0), which evaluates the \textit{stability} of detection performance across classes. Alongside PSDS scores, we calculate segment- and event-based F1 scores as an auxiliary metric. For F1 scores, we find optimal class-wise thresholds that result in the highest class-wise F1 scores and then macro-average across classes for a more fair comparison. For segment-based F1 scores, we use a segment length of 0.2s, and for event-based F1 scores, we use an onset and offset collar of 0.5s. For sample-level classification, we use the standard macro-averaged receiver operating characteristic area-under-curve (mROC) and average precision (mAP) metrics. 

\begin{table}[]
\setlength{\tabcolsep}{0.55em}
\centering
\begin{tabular}{lcccc}
\hline
Class (PSDS$_1$)           & Audio          & Vibration      & A+V   & A+V (Pretrain)          \\ \hline
Engine Knock            & .6702          & .3012          & .6480 & \textbf{.6765} \\
Belt Squeal      & .3975          & .2890          & .4126 & \textbf{.4665} \\
Exhaust Noise          & .7538          & .3347          & \textbf{.7567} & .7332 \\
Unstable Idle    & .0839          & .3353 & .2761 & \textbf{.3708}          \\
Internal Tick    & \textbf{.4751} & .3340          & .4075 & .4712          \\
Ambiguous Tick   & .4260           & .3062          & 4523  & \textbf{.4865} \\
Accessory Noise       & .4089          & .2029          & \textbf{.4127} & .3750 \\
Engine Rattle    & .3818 & .2763          & .3465 & \textbf{.4192}          \\
Startup Rattle   & .6628          & .5666          & .6594 & \textbf{.7179} \\
Trouble Starting & .7763          & .7441          & \textbf{.8354} & .8071 \\ \hline
\end{tabular}
\vspace{-0.5em}
\caption{Class-wise PSDS$_1$ scores. ``A+V": Audio+Vibration, ``A+V (Pretrain)": Audio+Vibration with pretraining and finetuning.}
\vspace{-1.5em}
\label{tab:classwise}
\end{table}

\vspace{-1.0em}
\subsection{Quantitative Results}
\label{ssec:quantresults}

As shown in Table \ref{tab:results}, we ablate our proposed method to investigate each modality's respective contribution to engine fault SED performance. To do so, we remove the other respective modality by excluding the CNN feature extractors $f_a$ and $f_v$, while keeping the rest of the network the same. The audio-only model becomes $\sigma(f_C(f_{GRU}(f_a(X_a))))$, while vibration-only becomes $\sigma(f_C(f_{GRU}(f_v(X_v))))$ when comparing to (\ref{eq:model}). We train these ablated models using the same implementation details in \ref{ssec:implementdetails}. When comparing audio-only to vibration-only, we see that the audio modality outperforms vibration across all metrics, showing that audio is a strong signal for engine fault SED. However, the vibration modality still significantly outperforms random predictions, showing there are still useful features being learned from vibration for SED. When comparing against our proposed audio+vibration fusion model, we see it outperforms any single modality across all metrics, showing that although we perform \textit{sound} event detection from audio, the fusion of vibration provides complementary information that improves SED performance. Specifically, we see a $2.5\%$ improvement in PSDS scores, $1\%$ improvement in F1 scores, and a $1.5\%$ improvement in sample-level scores. Further in Table \ref{tab:results} we show the performance of pretraining the fusion model with different $\lambda_1$ and $\lambda_2$ values, from \ref{ssec:pretraining}. When using only the classification loss term, $\lambda_1{=}1.0,\lambda_2{=}0.0$, we see that finetuning performance outperforms the fusion model without pretraining across most metrics, showing a simple classification loss on weakly-labeled data is useful for the final SED task. When using only the contrastive loss term, $\lambda_1{=}0.0,\lambda_2{=}1.0$, we see finetuning performance similar to the fusion model without pretraining, showing the contrastive loss alone does not create a strong model initialization for finetuning. When setting $\lambda_1{=}1.0$ and using various $\lambda_2$ values, we see a significant improvement in finetuning performance over all the non-pretrained and classification-only pretrained models. Specifically, we see this pretraining strategy outperform the non-pretrained fusion model by about $4.5\%$ on PSDS scores, $2.5\%$ on F1 scores, and $2.5\%$ on sample-level classification scores. 

In Table \ref{tab:classwise} we show class-wise SED performance to investigate each engine fault individually. As shown, the pretrained fusion model outperforms other methods across a majority of engine faults, showing both vibration fusion and the pretraining strategy improves SED performance. When comparing audio- and vibration-only with results from Table \ref{tab:results}, we see audio outperforming vibration, however we see vibration outperforming audio for certain engine faults. For example, unstable idle is better detected using vibration signals, as an unstable idle event often results in a non-audible shaking vibration of a vehicle. For classes like engine knock and accessory noise, we see that audio outperforms vibration as these engine faults are often only audible and do not cause significant abnormal vibrations. When we fuse audio and vibration, we see an improvement across most engine fault types, showing that there are still complementary features between the modalities that improve SED performance.

\begin{figure}
  \centering
  \includegraphics[width=1.0\linewidth]{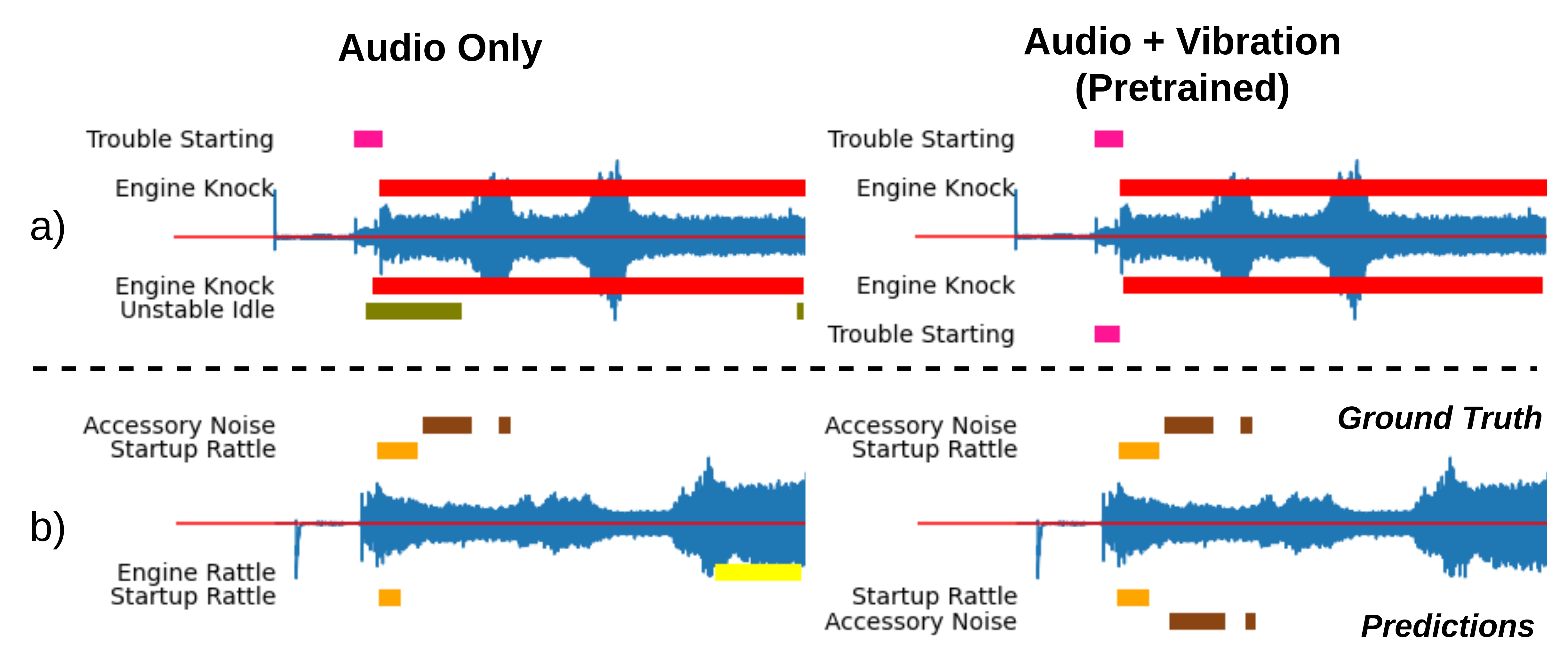}
  \vspace{-2.5em}
  \caption{Example engine fault detections. Events above the red line are ground truth labels and below the red line are predicted events.}
  \label{fig:positivequal}
  \vspace{-0.5em}
\end{figure}

\begin{figure}
  \centering
  \includegraphics[width=1.0\linewidth]{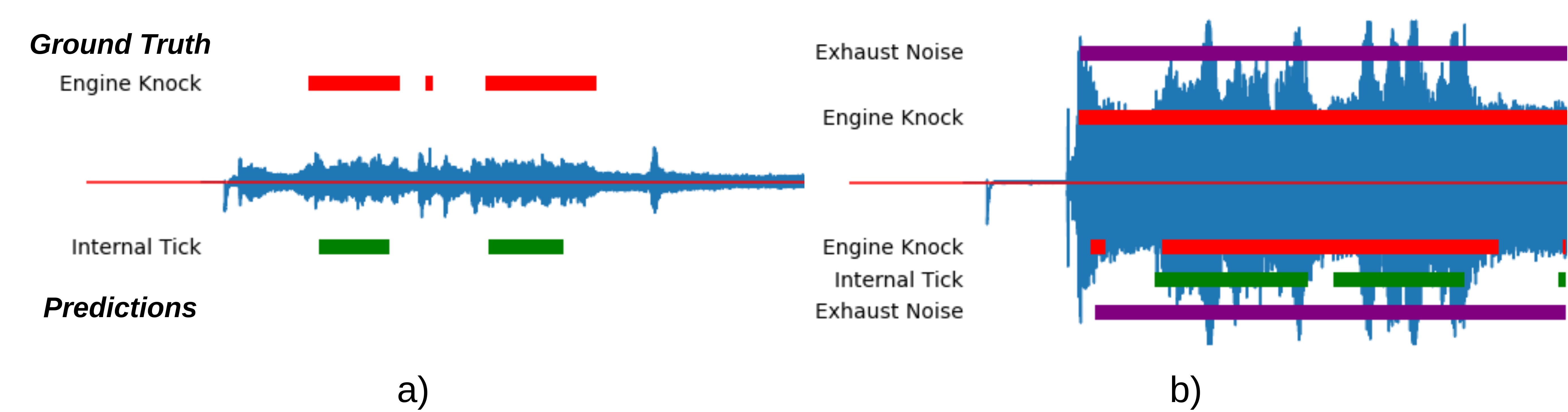}
  \vspace{-2em}
  \caption{Example failure cases of detecting engine fault sound events.}
  \label{fig:negativequal}
  \vspace{-1.5em}
\end{figure}

\vspace{-0.5em}
\subsection{Qualitative Results}
\label{ssec:qualresults}

In Figure \ref{fig:positivequal}, we show example detections of faulty engines. In Fig. \ref{fig:positivequal}a, we see that the audio model successfully detects the engine knock event, but misses the small-duration trouble staring event and has false positives on other faults. When combining the vibration modality with our pretraining strategy, we are able to remove the unstable idle false positive and more accurately detecting the trouble starting event. In Fig. \ref{fig:positivequal}b, we see the audio model captures the small-duration startup rattle, but misses the accessory noise and also has false positives. In the fusion model, we see all engine faults are detected with accurate onset and offset times. 

In Figure \ref{fig:negativequal}, we show example failure cases of our proposed SED model. In Fig. \ref{fig:negativequal}a, we see that the ground truth engine knock sound events are falsely-detected as internal tick events. These sound events are extremely similar sounds coming from similar areas of the engine, often confused even by engine experts. In Fig. \ref{fig:negativequal}b, we see the model is able to effectively capture the exhaust noise, but has false positives on internal tick and has poor onset and offset times of the knock event. This example vehicle has multiple significant engine faults that results in a very noisy audio and vibration recording, making the accurate temporal detection of the engine faults difficult. Further, this example shows the  \textit{polyphonic} nature of engine faults, that is, multiple sound events may be simultaneously occuring, adding to the difficulty of accurately detecting these events. 

\vspace{-0.5em}
\section{Conclusion}
\label{sec:conclusion}
\vspace{-0.5em}

In this paper, we explore engine fault sound event detection and show we are able to temporally detect fine-grained engine fault sound events using audio and vibration recordings across a wide variety of vehicles. Along with our collected data, we propose a simple multimodal CRNN-based SED architecture with a weakly-labeled pretraining strategy to perform engine fault SED. This work can be used to create automatic engine repair estimates, help mechanics diagnose engines, and serve as the basis for real-time engine condition monitoring. In the future we hope to explore other engine faults and more advanced SED architectures, like transformer-based methods.

\vspace{-0.5em}
\section{Acknowledgments}
\label{sec:acknowledgements}

This work was supported by ACV Auctions, Center for Identification Technology Research (CITeR), and National Science Foundation (NSF) under grant \#1822190 and partially under \#2229873. 
\bibliographystyle{IEEEbib}
\bibliography{strings,refs}

\end{document}